\documentclass[fleqn,twoside,twocolumn,nofootinbib,showkeys]{revtex4} 
\usepackage[sec,nocpr]{ujp} 

\begin{document}
\title[Magnetic Properties of Quantum Rings]
{MAGNETIC PROPERTIES OF QUANTUM\\ RINGS IN THE PRESENCE OF
SPIN-ORBIT\\ AND ELECTRON-ELECTRON INTERACTIONS}%
\author{O.S. Bauzha}
\affiliation{Taras Shevchenko National University of Kyiv}
\address{64, Volodymyrs'ka Str., Kyiv 01033, Ukraine}
\email{asb@univ.kiev.ua}%
\udk{621.38}  \razd{\secviii}

\autorcol{O.\hspace*{0.7mm}Bauzha}

\setcounter{page}{888}%

\begin{abstract}
The separate and combined influences of the spin-orbit and
electron-electron interactions on the electron magnetization in
quantum rings have been studied theoretically on the basis of the
spin-density-functional theory and the Kohn--Sham equation used for
the calculation of electron states in two-dimensional parabolic
quantum rings containing from two to six electrons. The
magnetization of electrons in a quantum ring is calculated at
zero temperature. The revealed abrupt changes in the ring
magnetization are associated with the crossing of electron states
that occurs if the spin-orbit and/or electron-electron interactions
are taken into consideration.
\end{abstract}
 \maketitle

\section{Introduction}

Spintronics is a branch of electronics, in which the principle of device
operation is based on the use of electron's spin degree of freedom. Some
devices belonging to this group have already proved their importance even at
the commercial level. In particular, spin-valve reading heads of hard disks
can serve as an example of such an application. The review of this direction
can be found in work \cite{1}.

This research is devoted to quantum rings (QRs). The study of those
systems is challenging, because both QRs and quantum dots (QDs) can
be regarded as artificial atoms. In contrast to ordinary atomic
systems, QRs possess a variety of features including the possibility
to manipulate their magnetic properties with the use of an external
electric field \cite{2, 3}. Quantum rings can be used as qubits in
quantum computers. Therefore, their properties studied in this work
should be taken into consideration when designing such devices.

\begin{figure*}
\vskip1mm
\includegraphics[width=3.5cm]{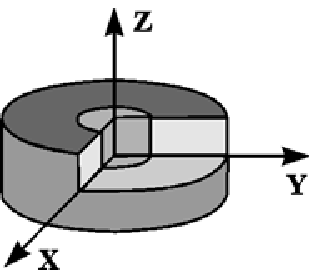}\hspace*{1.5cm}\includegraphics[width=3.0cm]{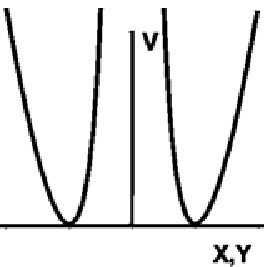}\hspace*{1.5cm}\includegraphics[width=3.5cm]{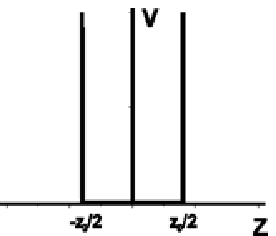}\\
{\it a\hspace*{5.0cm}b\hspace*{5.0cm}c} \vskip-3mm\caption{Schematic
images of a quantum ring and the corresponding potential profiles
}
\end{figure*}

The aim of this work is to analyze the influence of the spin-orbit interaction in
weak magnetic fields on the magnetic properties of small semiconducting QRs.
The magnetization of QRs in a quasiparabolic potential for electrons was
calculated with regard for for the spin-orbit (SO) and electron-electron (EE)
interactions. In the previous works, theoretical researches of the magnetic
properties of QDs with regard for the SO \cite{4} and EE \cite{5}
couplings were carried out. The magnetic properties of QRs taking the SO
interaction into account were also calculated \cite{6}, but the EE coupling
for QRs was not considered.

To describe the influence of spin-orbit interaction on QR magnetization, the
Rashba approximation was applied \cite{7,8}. While calculating the influence
of EE coupling in quantum-sized structures, the Hartree and Hartree--Fock
approximations, as well as the theory of spin density functional are used
\cite{9, 10, 11, 12, 13}. In work \cite{14}, various approaches to the
calculation of electron ground state energies in QDs were compared in detail
(the cases of filling a QD with two to thirteen electrons were examined). In
particular, it was shown that the calculation error for the energies of
electron ground states obtained in the framework of the spin-density
functional theory does not exceed 2.5\%. Therefore, the mentioned theory is
also applied in this work to describe the EE interaction in structures of the
same class.

\section{Method}

To describe the properties of magnetic rings, a simple model
potential for an insulated two-dimensional ring located in the plane
$XY$ was used \cite{15, 16} (see Fig.~1),
\begin{equation}
V_{c}\left(  r\right)  =\frac{a_{1}}{r^{2}}+a_{2}r^{2}-V_{0},\label{eq1}%
\end{equation}
where $V_{0}=2\sqrt{a_{1}a_{2}}$. In this model, both the ring
radius and width can be selected independently. In the previous works
\cite{15, 16}, the energy spectrum and the magnetization at zero
temperature were calculated, as well as the wave functions in a
constant magnetic field applied perpendicularly to the ring plane.
Neither the SO, nor the EE interaction was taken into consideration
in those works.

Potential (\ref{eq1}) has the following properties:

(a)~there exists a minimum $V\left(  {r_{0}}\right)=0$ at
\begin{equation}
r=r_{0}=\left(\!  {\frac{a_{1}}{a_{2}}}\!\right)  ^{\!{1}/{4}}\!, \label{eq2}%
\end{equation}
which determines the average ring radius;

(b)~at $r\approx r_{0}$, the potential has a simple parabolic shape
$V\left(  r\right)  \approx\frac{1}{2}m\omega_{0}^{2}\left(
{r-r_{0}}\right) ^{2}$ (Fig.~2), where the parameter
$\omega_{0}=\sqrt{\frac{8a_{2}}{m}}$ characterizes the size of
the potential well, and $m$ is the effective electron mass.

Potential (\ref{eq1}) can also be used to describe a number of other physical
systems, in particular, a 1D ring (at $r_{0}=\mathrm{const}$ and $\omega
_{0}\rightarrow\infty$), a 2D straight wire (at $\omega_{0}=\mathrm{const}$
and $r_{0}\rightarrow\infty$), a quantum dot (at $a_{1}=0$), an isolated
antidot (at $a_{2}=0$), and others.

\begin{figure}%
\includegraphics[width=7.5cm]{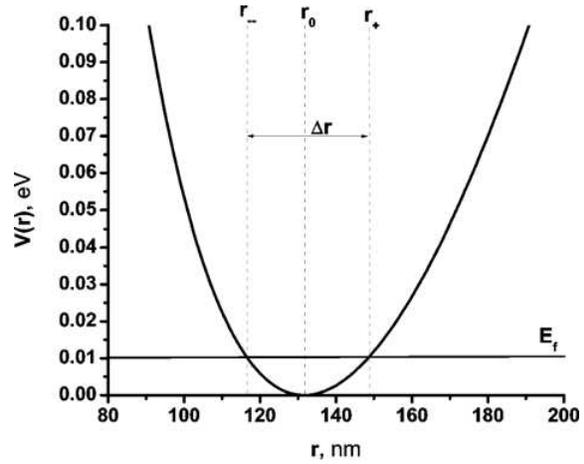}
\vskip-3mm\caption{Profile of the quantum-ring potential  }
\end{figure}

\subsection{One-electron Hamiltonian}

Provided that a uniform magnetic field $\mathbf{B}$ is applied along the
QR symmetry axis (the $z$-axis), the one-electron Hamiltonian can be written
down in the polar coordinates ${\{}r,\varphi{\}}$ \cite{17} as follows:
\[
 H_1 =-\frac{\hbar ^2}{2m(E)}\left[ {\frac{\partial }{r\partial
r}r\frac{\partial }{\partial r}+\frac{1}{r^2}\frac{\partial
^2}{\partial \phi ^2}} \right]-\frac{i}{2}m\omega _c (E,B)\,\times
\]\vspace*{-7mm}
\[\times\,\frac{\partial }{\partial \phi } +\frac{1}{8}m(E)\omega _c^2 (E,B)r^2+V_c (r)+V_{so}^{\rm R} (r,\phi )\,+\]\vspace*{-7mm}
\begin{equation}
\label{eq3} +\,\frac{1}{2}\sigma _z \mu _{\rm B} g(E)B.
\end{equation}
Here, the first term is responsible for the kinetic energy of the
electron, the second and third terms describe the influence of a
magnetic field on the electron motion, $V_{c}(r)$ is the confining
potential, $V_{so}^{\rm R}(r,\phi)$ the energy of spin-orbit
interaction, and the last term describes the interaction between the
electron spin and the magnetic field. For the effective mass, we use
the expression \cite{8, 18}\vspace*{-2mm}
\begin{equation}
\frac{1}{m(E)}\!=\!\frac{1}{m(0)}\frac{E_{g}\left(
{E_{g}\!+\Delta}\right) }{\left(  {3E_{g}\!+2\Delta}\right)
}\left[\! {\frac{2}{E+E_{g}}+\frac
{1}{E+E_{g}\!+\Delta}}\!\right]\!, \label{eq4}%
\end{equation}
where $E$ and $m(E)$ are the electron energy and mass, respectively, in the
conduction band; $m(0)$ is the effective electron mass near the conduction
band bottom; $E_{g}$ the energy gap width; and $\Delta$ the spin-orbit
splitting of the valence band. In formula (\ref{eq3}), the quantity
\[
\omega_{c}(E,B)=\frac{eB}{m(E)}%
\]
is the electron cyclotron frequency, and $\sigma_{z}$ is the Pauli $z$-matrix. For
the $g$-factor, we used the \mbox{expression \cite{19}}%
\begin{equation}
g(E)=2\left[
{1-\frac{m_{0}}{m(E)}\frac{\Delta}{3(E_{g}+E)+2\Delta}}\right]\!, \label{eq5}%
\end{equation}
$\mu_{\rm B}=e\hbar/2m_{0}$ is the Bohr magneton, $e$ the electron
charge, and $m_{0}$ the free electron mass.

The operator of SO interaction energy, which was introduced
by Rashba, is taken in the form \mbox{\cite{8, 20,21,22}}
\begin{equation}
V_{so}^{\rm R}(r,\phi)=\sigma_{z}\alpha\frac{dV_{c}(r)}{dr}\left( \! {k_{\phi}%
+\frac{e}{2\hbar}Br}\!\right)\!, \label{eq6}%
\end{equation}
where $k_{\phi}=-i(1/r)\partial/\partial\phi$, and $\alpha$ is the parameter
of the spin-orbit interaction introduced by Rashba \cite{8}.

The stationary Schr\"{o}dinger equation with Hamiltonian (\ref{eq3}) has no
analytical solution, so it was solved numerically. The results of
corresponding calculations and the electron wave function are
presented in work \cite{6}.

\subsection{Spin density functional theory}

While calculating the energies and the wave functions of electrons confined in a
certain region, electrostatic and electromagnetic interactions between
particles have to be taken into account. The exact solution of this problem is
extremely difficult, and a number of simplifications have to be applied. The
density functional theory forms a basis for the description of electrons in a
confining potential. This theory allows one to give an equivalent one-particle
formulation for the complicated many-particle problem.

When calculating the electron spectra for a two-dimensional quasiparabolic
quantum ring taking the EE and SO interactions into account simultaneously,
the corresponding Kohn--Sham equation was solved self-consistently
\cite{9,11},
\[ \left[ {H_1 +\frac{e^2}{\kappa }\int {\frac{w({\rm {\bf
{r}'}})}{\left| {{\rm {\bf r}}-{\rm {\bf {r}'}}} \right|}}
d{r}'+\frac{\delta E_{xc} \left( {w,\zeta } \right)}{\delta w^\sigma
({\rm {\bf r}})}} \right] \psi _{n,l}^\sigma \left( {\rm {\bf r}}
\right)=\]
\begin{equation}
\label{eq7}= \varepsilon _{n,l}^\sigma \psi _{n,l}^\sigma \left(
{\rm {\bf r}} \right)\!,
\end{equation}\vspace*{-5mm}
\begin{equation}
w(\mathrm{\mathbf{r}})=\sum\limits_{\sigma}{w^{\sigma}(\mathrm{\mathbf{r}}%
)}=\sum\limits_{\sigma}{\sum\limits_{n,l}{\left\vert {\psi_{n,l}^{\sigma
}(\mathrm{\mathbf{r}})}\right\vert ^{2}}},\label{eq8}%
\end{equation}
where $H_{1}$ is the electron Hamiltonian in the one-electron approximation
(Eq.~(\ref{eq3})). The superscript $\sigma$ corresponds to the electron spin,
$\zeta(\mathrm{\mathbf{r}})$ is the local spin polarization, $\kappa$ the
dielectric constant, and $E_{xc}$ the functional of exchange-correlation
energy, which was used in the local density approximation \cite{10}.
Hereafter, the atomic units are used, the radius is reckoned in terms of
the effective Bohr radius ($\kappa\hbar^{2}/m^{\ast}e^{2}),$ and the energy is taken in
effective Hartree units ($m^{\ast}e^{4}/\kappa^{2}\hbar^{2})$.
\begin{equation}
E_{xc}=\int{w(\mathrm{\mathbf{r}})\varepsilon_{xc}\left[
{w(\mathrm{\mathbf{r}}),\zeta(\mathrm{\mathbf{r}})}\right]  }%
d\mathrm{\mathbf{r}},\label{eq9}%
\end{equation}\vspace*{-5mm}
\begin{equation}
\zeta(\mathrm{\mathbf{r}})=\frac{w^{\uparrow}(\mathrm{\mathbf{r}%
})-w^{\downarrow}(\mathrm{\mathbf{r}})}{w(\mathrm{\mathbf{r}})},\label{eq10}%
\end{equation}
where $\varepsilon_{xc}\left[  {w(\mathbf{r}),\zeta(\mathbf{r})}\right]  $ is
the exchange-correlation energy per one particle in a uniform spin-polarized
gas, which is considered to be a sum of the exchange and correlation energies
\cite{10},
\begin{equation}
\varepsilon_{xc}\left[  {w(\mathrm{\mathbf{r}}),\zeta(\mathrm{\mathbf{r}}%
)}\right]  =\varepsilon_{x}\left[  {w(\mathrm{\mathbf{r}}),\zeta
(\mathrm{\mathbf{r}})}\right]  +\varepsilon_{c}\left[  {w(\mathrm{\mathbf{r}%
}),\zeta(\mathrm{\mathbf{r}})}\right]\!.\label{eq11}%
\end{equation}
Below, while considering the electron-electron interaction, only the
exchange energy component will be taken into consideration. In the case of
two-dimensional electron gas, the exchange interaction looks like
\begin{equation}
\label{eq12} \varepsilon _x \left[ {w,\zeta }
\right]=-\frac{4}{3r_{\rm B} }\sqrt {\frac{2w}{\pi }} \left[ {\left(
{1+\zeta } \right)^{3/2}+\left( {1-\zeta } \right)^{3/2}} \right]\!,
\end{equation}
where $r_{\rm B}$ is the Bohr radius. Then, the ground-state energy
of a quantum ring with $N$ electrons has the form
\[
 E_{\rm tot} (N)=\sum\limits_{n,l,\sigma } {\varepsilon _{n,l}^\sigma +}
\frac{e^2}{2\kappa }\int {\frac{w({\rm {\bf r}})w({\rm {\bf
{r}'}})}{\left| {{\rm {\bf r}}-{\rm {\bf {r}'}}} \right|}} d{\rm
{\bf r}}d{\rm {\bf {r}'}}-
\]\vspace*{-5mm}
\begin{equation}
\label{eq13}
 -\sum\limits_\sigma {\int {w^\sigma ({\rm {\bf r}})} } \frac{\delta E_{xc}
\left( {w,\zeta } \right)}{\delta w^\sigma ({\rm {\bf r}})}d{\rm
{\bf r}}+E_{xc}.
\end{equation}
At zero temperature, the magnetization is determined as follows:%
\begin{equation}
M=-\frac{\partial E_{\rm tot}}{\partial B}.\label{eq14}%
\end{equation}

\begin{figure*}
\vskip1mm
\includegraphics[width=5cm]{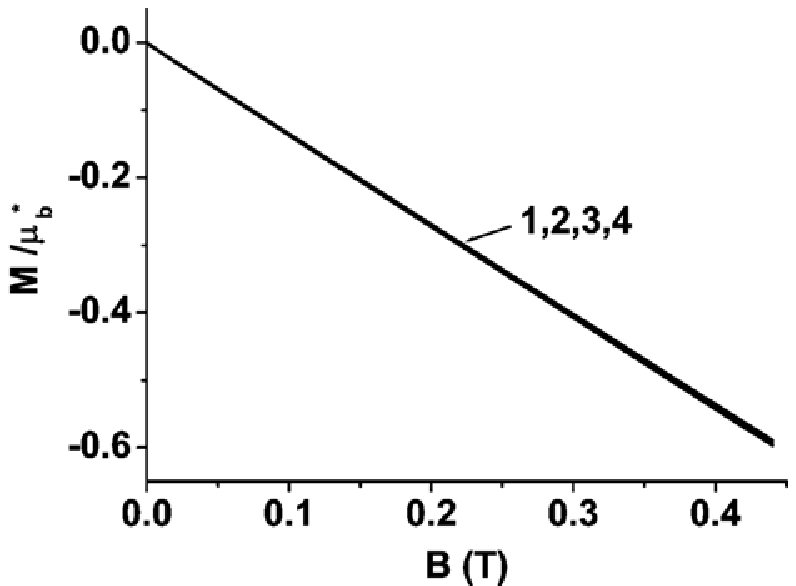}\hspace*{0.5cm}\includegraphics[width=5cm]{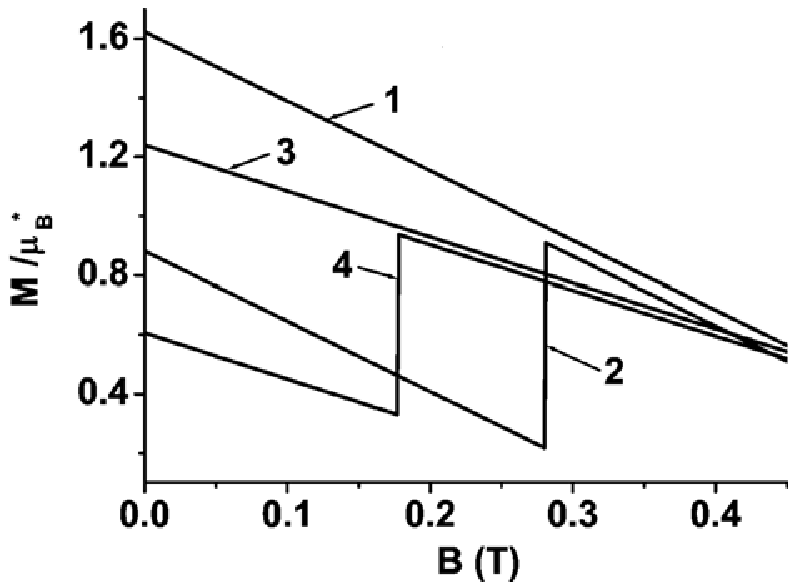}\hspace*{0.5cm}\includegraphics[width=5cm]{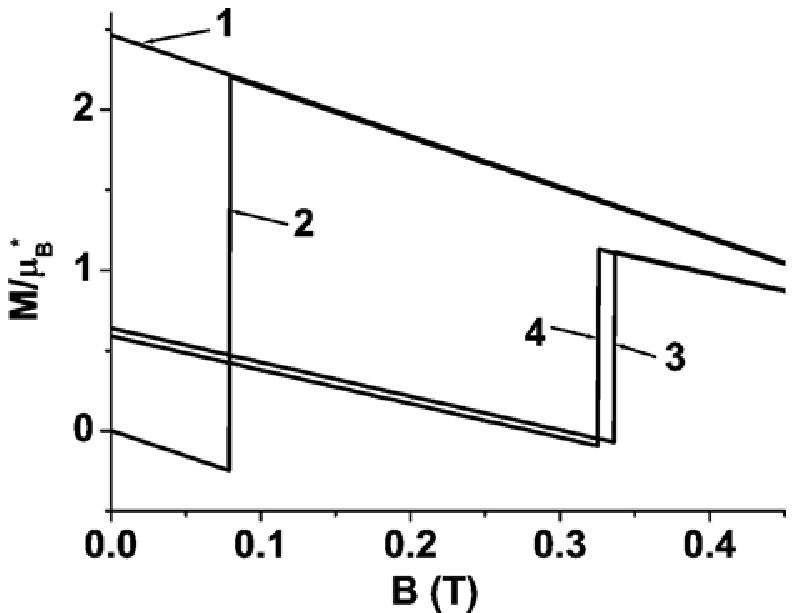}\\
{\it a\hspace*{5.25cm}b\hspace*{5.25cm}c}\\[3mm]
\hspace*{0.5cm}\includegraphics[width=5cm]{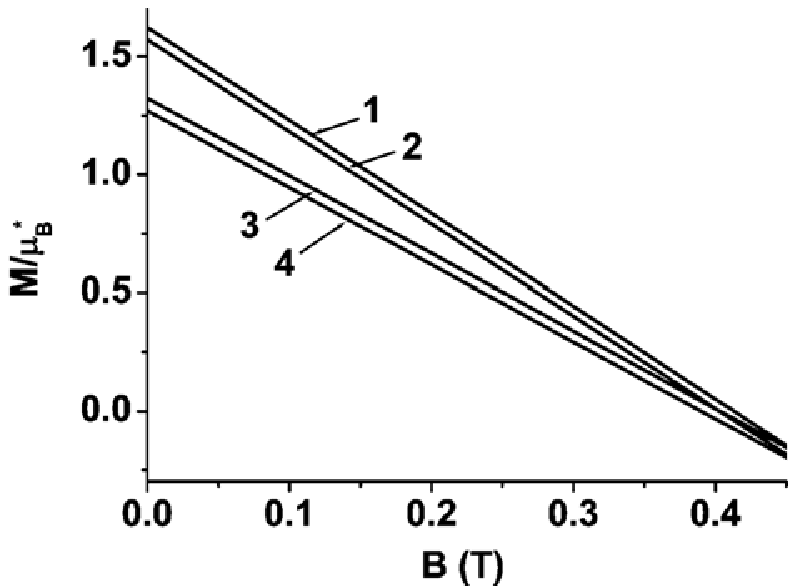}\hspace*{0.5cm}\includegraphics[width=5cm]{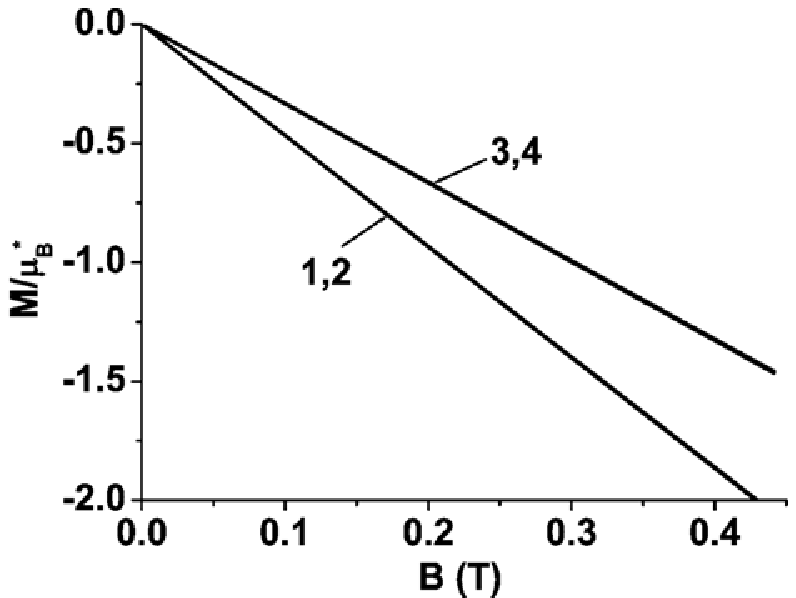}\\
{\it d\hspace*{5.25cm}e}
 \vskip-3mm\caption{Dependences of the
magnetization in the InSb quantum ring filled with two to six
electrons (panels \textit{a} to \textit{e}, respectively) on the
magnetic applied field. The ring dimensions are
$r_{0}\approx26$\textrm{~nm} and $\Delta r\approx18$\textrm{~nm.}
Calculations were carried out taking no SO and EE interactions into
account (curves~\textit{1}), taking only SO interaction into account
(curves~\textit{2}), taking only EE interaction into account
(curves~\textit{3}), and taking both SO and EE interactions into
account (curves~\textit{4}). $\mu_{\rm B}^{\ast}=e\hbar/2m(0)$
}\vspace*{-1.5mm}
\end{figure*}

The solution of Eq.~(\ref{eq7}) was obtained in the framework of the
self-consistent field approach. Namely, the energy of an electron
was determined assuming that the states of other electrons were
known. Then, the obtained solution was used to correct the states of
other electrons and the potential created by them. While carrying
out self-consistent calculations, the Broyden algorithm \cite{23}
was used.

\section{Results of Calculations}

The influence of spin-orbit and electron-electron interactions on the ring
magnetization (ring's magnetic moment) at zero temperature and a small
number of electrons in the quantum ring was studied theoretically. For an InSb
quantum ring, the following parameters were selected \cite{17, 18, 24}:
$m(0)=0.014m_{0}$, $E_{g}=0.24~\mathrm{eV}$, $\Delta=0.81~\mathrm{eV}$, and
$\alpha=5~\mathrm{nm}^{2}$. Consider quantum rings with the following widths
$\Delta r$ and average radii $r_{0}$:

(a)~$r_{0}\approx26$\textrm{~nm}, $\Delta r\approx18$\textrm{~nm}
($a_{1}=31.5~\mathrm{eV (nm)}^{2}$, $a_{2}=0.094~\mathrm{meV}%
/\mathrm{nm}^{2}$;

(b) $r_{0}\approx24$\textrm{~nm}, $\Delta r\approx18$\textrm{~nm}
($a_{1}=67~\mathrm{eV (nm)}^{2}$, $a_{2}=0.145~\mathrm{meV}/\mathrm{nm}%
^{2})$;

(c) $r_{0}\approx26$\textrm{~nm}, $\Delta r\approx16$\textrm{~nm}
($a_{1}=43~\mathrm{eV (nm)}^{2}$, $a_{2}=0.094~\mathrm{meV}/\mathrm{nm}%
^{2})$.

 The parameters $a_{1}$ and $a_{2}$ (see Eq.~(\ref{eq1})) were so
selected to trace the influence of small changes in the average
radius (item~b) and the width (item~c) of a QR on the magnetic
properties of rings. The QR sizes selected for calculations
correspond to the typical dimensions of structures that are grown
up experimentally \cite{25}.

\begin{figure*}
\vskip1mm
\includegraphics[width=5cm]{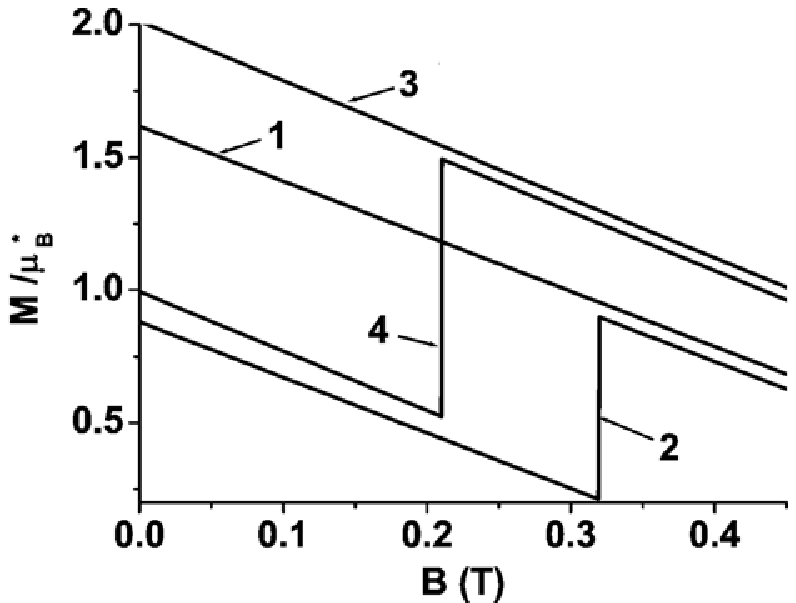}\hspace*{0.5cm}\includegraphics[width=5cm]{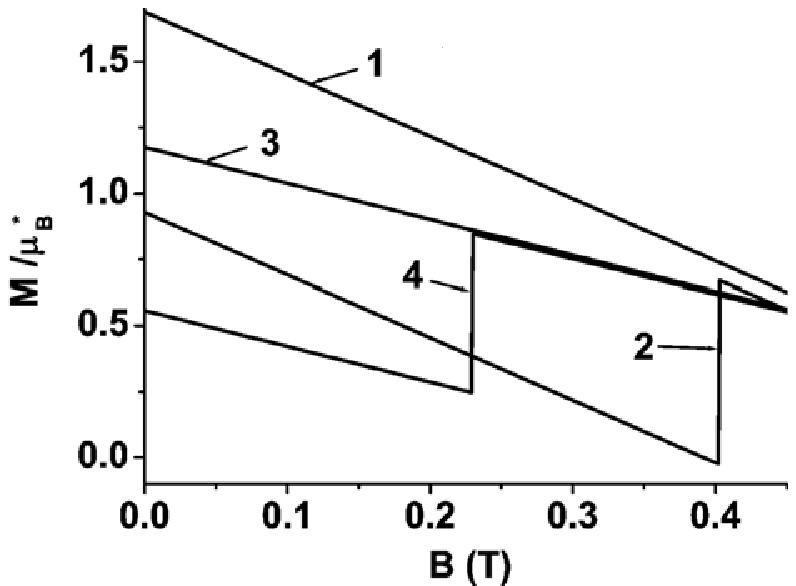}\\
{\it a\hspace*{5.25cm}b} \vskip-3mm\caption{ Dependences of the
magnetization in an InSb quantum ring filled with three electrons
on the magnetic applied field. The ring dimensions are
$r_{0}\approx24$\textrm{~nm} and $\Delta r\approx18$\textrm{~nm}
(panel \textit{a}) and $r_{0}\approx26$\textrm{~nm} and $\Delta
r\approx 16$\textrm{~nm} (panel \textit{a})\textrm{.} The curve
notation is the same as in Fig.~3 }\vskip3mm
\end{figure*}

\begin{figure*}
\includegraphics[width=5cm]{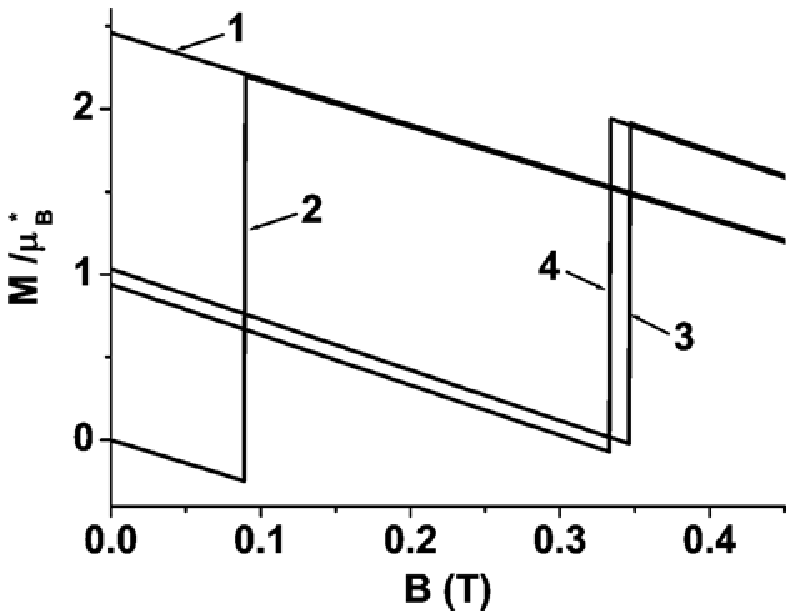}\hspace*{0.5cm}\includegraphics[width=5cm]{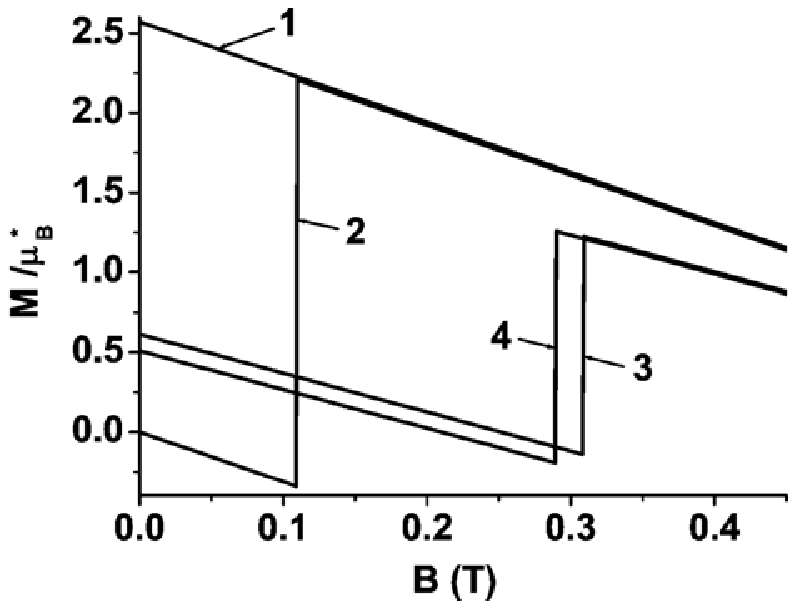}\\
{\it a\hspace*{5.25cm}b} \vskip-3mm\caption{The same as in Fig.~4,
but for quantum rings filled with four electrons  }\vspace*{1mm}
\end{figure*}

The calculated magnetizations of quantum rings with two to six electrons are
illustrated in Figs.~3 to 5. For the sake of comparison, the magnetizations
calculated for rings with the same number of electrons, but
taking no account of spin-orbit or electron-electron interaction, are also
depicted in the relevant figures.

First, consider the quantum ring (a) with the dimensions $r_{0}\approx
26$\textrm{~nm} and $\Delta r\approx18$\textrm{~nm}. From
Figs.~3,~\textit{a}, \textit{d}, and \textit{e}, one can see that, in
the case where the quantum ring is filled with two, five, or six
electrons, the influence of the spin-orbit interaction is almost
unnoticeable. The electron-electron interaction gives rise to a
variation in the slope of the magnetization dependence on the
magnetic field. If the QR is filled with two (Fig.~3,~\textit{a})
and six (Fig.~3,~\textit{e}) electrons, the quantum ring reveals
diamagnetic properties. This behavior corresponds to the situation
when the external electron shells are filled.

As one can see from Fig.~3,~\textit{c}, if the ring is filled with
four electrons, the account of the electron-electron interaction brings
about the emergence of a jump in the magnetization
(curve~\textit{2}). If the ring is filled with four electrons, the
account of the electron-electron interaction also results in the
emergence of a jump in the magnetization, but at higher magnetic
fields (curve~\textit{3}). With regard for both
electron-electron and spin-orbit interactions (Fig.~3,~\textit{c},
curve~\textit{4}) shifts the magnetization jump toward lower
fields. At the same time, the electron-electron interaction alone
does not induce the emergence of a magnetization jump
(Fig.~3,~\textit{d}, curve~\textit{3}). The latter arises when the
spin-orbit interaction is also taken into account
(curves~\textit{2}~and~\textit{4}).

In Fig.~4, the dependences of the ring magnetization on the magnetic
field are exhibited for QR (b) with the dimensions
$r_{0}\approx24$\textrm{~nm} and $\Delta r\approx18\mathrm{~nm}$
(Fig.~4,~\textit{a}) and QR (c) with the dimensions
$r_{0}\approx26$\textrm{~nm} and $\Delta r\approx16\mathrm{~nm}$
(Fig.~4,~\textit{b}), the both filled with 3 electrons. In
Fig.~4,\textit{a}, the number of electrons in the ring is the same
as in Fig.~3,~\textit{b}, but the average ring radius is smaller.
The ring with the dependences shown in Fig.~4,~\textit{b} has a
smaller width in comparison with the ring corresponding to the
dependences in Fig.~2,~\textit{b}. One can see that a decrease of
either the average ring radius or the ring width results in the
enhancement of the influence of both the spin-orbit and
electron-electron interactions. The considered changes in the QR
sizes shift the magnetization jump toward higher magnetic fields
(Figs.~3,~\textit{b}; 4,~\textit{a}; and 4,~\textit{b}).

In Fig.~5, the dependences of the ring magnetization on the magnetic
field are exhibited for QR (b) with the dimensions
$r_{0}\approx24$\textrm{~nm} and $\Delta r\approx18\mathrm{~nm}$
(Fig.~5,~\textit{a}) and QR (c) with the dimensions $r_{0}\approx$
$\approx26$\textrm{~nm} and $\Delta r\approx16\mathrm{~nm}$
(Fig.~4,~\textit{b}), the both being filled with 4 electrons. The
figures also demonstrate that there is a shift in the positions of
the magnetization jumps. Namely, the jumps become shifted toward
higher magnetic fields. The distance between magnetization jumps
when both the SO and EE interactions are taken into account is
larger than if only the EE interaction is considered
(Figs.~3,~\textit{c}; 5,~\textit{a}; and 5,~\textit{b};
curves~\textit{3} and~\textit{4}).

\section{Conclusions}

Quantum rings with completely filled external electron shells~--
with two (Fig.~3,~\textit{a}) or six (Fig.~3,~\textit{e}) electrons
in the QR~-- behave like diamagnetics. The account of both SO and EE
interactions does not change the magnetic properties of QRs with
filled external shells. If the external electron shell of a QR is
not filled, there may emerge jumps in the dependence of the ring
magnetization on the applied magnetic field (Figs. 3,~\textit{b},
3,~\textit{c}, 4, and 5). Those jumps are the consequences of the SO
and EE interactions.

The influence of the SO and EE interactions results in the splitting of
energy levels in the zero magnetic field for the electrons in the
quantum ring. The splitting brings about the possibility of the crossing
between the energy levels if an external magnetic field is applied.
This is the factor that is responsible for the emergence of jumps in
the ring magnetization. If either or both of QR sizes~-- the average
radius and the width~-- diminishes, the influence of the SO and EE
interactions becomes stronger. Therefore, by changing the material
or the geometry of QRs, it is possible to control their magnetic
properties.

\vspace*{-5mm}

\rezume{%
О.С. Баужа}{МАГНІТНІ ВЛАСТИВОСТІ КВАНТОВИХ\\ КІЛЕЦЬ ПРИ ВРАХУВАННІ
СПІН-ОРБІТАЛЬНОЇ\\ ТА ЕЛЕКТРОН-ЕЛЕКТРОННОЇ ВЗАЄМОДІЇ} {У роботі
наведено теоретичний розгляд впливу спін-орбітальної (СО) та
електрон-електронної (ЕЕ) взаємодій на електронну намагніченість
квантових кілець (КК). У дослідженні використано теорію функціонала
густини та рівняння Кона--Шема для розрахунку енергетичних рівнів
електронів у дворозмірному квазіпараболічному квантовому кільці,
заповненому 2--6-ма електронами. Намагніченість електронів у
квантовому кільці розрахована при нульовій температурі. Різка зміна
намагніченості пов'язана з перетином енергетичних рівнів електронів
(ці перетини є наслідком врахування спін-орбітальної або
електрон-електронної взаємодії).}

\end{document}